\documentclass[10pt, conference]{IEEEtran}
\ifCLASSINFOpdf
   \usepackage[pdftex]{graphicx}
   \graphicspath{{../pdf/}{../jpeg/}}
   \DeclareGraphicsExtensions{.pdf,.jpeg,.png}
\else
   \usepackage[dvips]{graphicx}
   \graphicspath{{../eps/}}
   \DeclareGraphicsExtensions{.eps}
\fi

\usepackage{cite}
\usepackage{amsmath,amssymb,amsfonts}
\usepackage{algorithmic}
\usepackage[ruled]{algorithm2e}
\usepackage{graphicx}
\usepackage{textcomp}

\usepackage{booktabs} 
\usepackage{color}
\usepackage{multirow}
\usepackage{subfigure}

\usepackage[T1]{fontenc}
\usepackage{fix-cm}

\def\BibTeX{{\rm B\kern-.05em{\sc i\kern-.025em b}\kern-.08em
    T\kern-.1667em\lower.7ex\hbox{E}\kern-.125emX}}

\let\labelindent\relax

\usepackage{enumitem}
\setlist[enumerate]{itemsep=0mm}

\usepackage{color}
\usepackage{subfigure}
\graphicspath{{figures/}{pictures/}{images/}{./}} 

\usepackage{multirow}
\usepackage{array}

\usepackage[table]{xcolor}

\newtheorem{definition}{Definition}[section] 

\begin{document}

\title{Predicting Opioid Relapse Using Social Media Data}

\author{
    \IEEEauthorblockN{Zhou Yang}
    \IEEEauthorblockA{
        \textit{Department of Computer Science} \\
        \textit{Texas Tech University}\\
            Lubbock, Teas, USA \\
            zhou.yang@ttu.edu}
    \and
    \IEEEauthorblockN{Long Nguyen}
    \IEEEauthorblockA{
        \textit{Computer Science Department} \\
        \textit{Texas Tech University}\\
            Lubbock, Texas, USA \\
            long.nguyen@ttu.edu}
    \and
    \IEEEauthorblockN{Fang Jin}
    \IEEEauthorblockA{
        \textit{Computer Science Department} \\
        \textit{Texas Tech University}\\
            Lubbock, Texas, USA \\
            fang.jin@ttu.edu}
}

\maketitle

\begin{abstract}

Opioid addiction is a severe public health threat in the U.S, causing massive deaths and many social problems. Accurate relapse prediction is of practical importance for recovering patients since relapse prediction promotes timely relapse preventions that help patients stay clean. In this paper, we introduce a Generative Adversarial Networks (GAN) model to predict the addiction relapses based on sentiment images and social influences. Experimental results on real social media data from Reddit.com demonstrate that the GAN model delivers a better performance than comparable alternative techniques. The sentiment images generated by the model show that relapse is closely connected with two emotions `joy' and `negative'. This work is one of the first attempts to predict relapses using massive social media data and generative adversarial nets. The proposed method, combined with knowledge of social media mining, has the potential to revolutionize the practice of opioid addiction prevention and treatment.
\end{abstract}


%
\IEEEpeerreviewmaketitle

\section{Introduction}\label{sec:Introduction}
\begin{figure}[tbp]
 \centering
 \includegraphics[width=\linewidth]{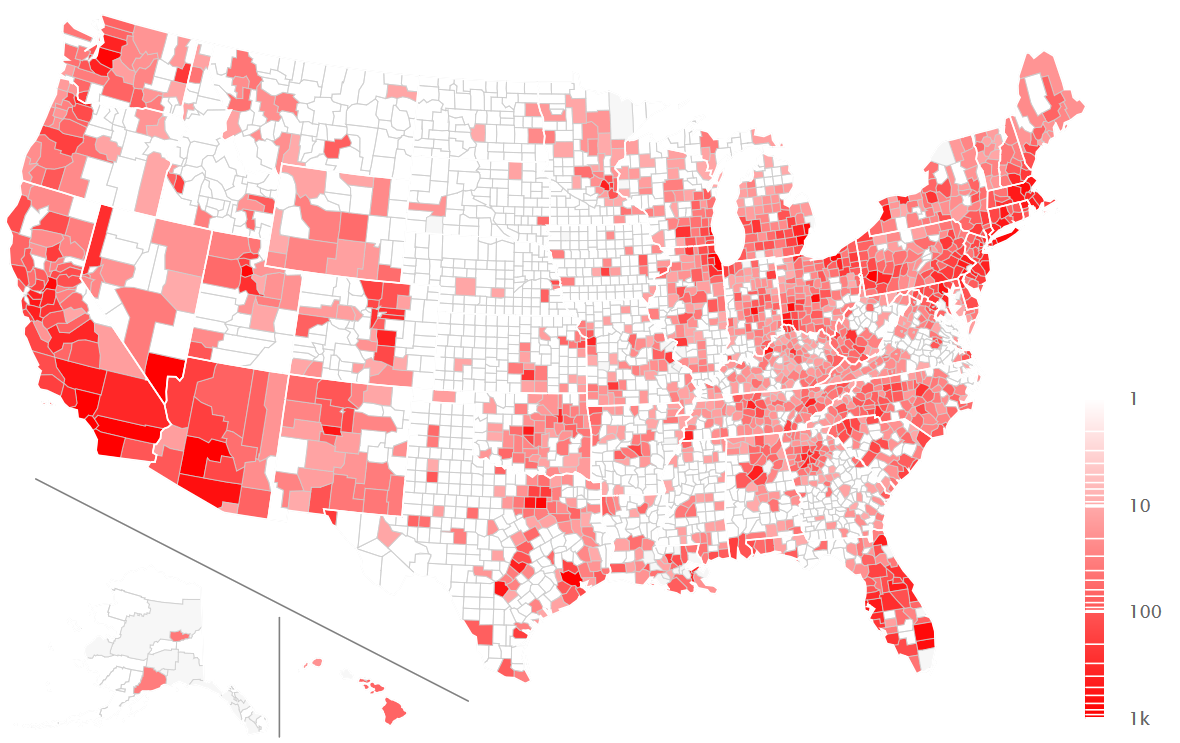}
 \caption{County level drug overdose deaths in the U.S.A 2017~\cite{Death_rates}.}
 \label{fig:heat_map}
\end{figure}

Opioid addiction is a severe public health threat in the U.S, causing massive deaths and many social problems~\cite{overdose_death}. According to the latest statistics of National Institute on Drug Abuse (NIDA, 2017), more than 115 Americans die after overdosing on opioids on a daily basis, and nearly 64,000 people died of drug overdoses in the US in 2016, the most lethal year of the drug overdose epidemic (NIDA, 2017). Moreover, millions of Americans have been influenced by opioid-related problems. It is estimated that 2.1 million people suffer from substance use disorders related to prescription opioid pain relievers in the United States alone~\cite{overdose_death}, where there has been a significant increase in opioid-related deaths from 2000 to 2017, and the death toll is still rising~\cite{Death_rates}. The opioid crisis has social impacts beyond the increased death toll. The status of the opioid crisis in the US is shown in Figure~\ref{fig:heat_map}; other consequences include a rise in the number of infants born dependent on opioids~\cite{patrick2015increasing, tolia2016increasing} as well as the spread of infectious diseases such as HIV and hepatitis C~\cite{conrad2015communfity}. 
Opioid epidemic has now deteriorated into a full-scale national pandemic, leading to national concern because of its negative impacts on health, social security and economics

\begin{figure}[t]
 \centering
 \includegraphics[width=\linewidth]{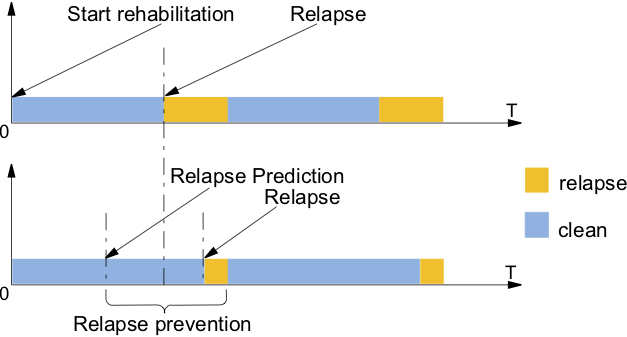}
 \caption{Rehabilitation process. Relapse is an inevitable part of the rehabilitation process, but accurate relapse prediction combined with other supports can postpone or prevent relapse, and hence compress the relapsing period~\cite{marlatt2005relapse, geddes2003relapse}.}
 \label{fig:relapsing}
\end{figure}

Effective ways to mitigate the opioid crisis are thus urgently needed. 
Most of the previous work in this area has tended to focus on discovering drug-related adverse events~\cite{bian2012towards}, for example by proposing platforms to identify prescription drug abuse~\cite{cameron2013predose}, monitor medication abuse via social media~\cite{sarker2016social}, or detect opioid addicts~\cite{miller2010using,fan2017social}. As yet, however, there have been few attempts at opioid addiction relapse prediction. The ability to predict relapses is critically important; as the data shown in Figure~\ref{fig:relapsing} indicate, since this could potentially provide substantial support for better addiction prevention and treatment.

The addiction relapse process involves three main phases, namely emotional, mental, and physical relapses~\cite{theraleighhouse,brownell1986understanding}. Relapse is a gradual process that could start days to weeks before the actual consumption of drugs. Typically, addiction relapse begins on an emotional level, which then goes on to become a physical relapse if no action is taken to support the recovering addict. These emotional states, once confirmed, can thus be regarded as an indicator for relapse prediction. An accurate relapse prediction is essential to address the opioid epidemic since a precise relapse prediction makes it possible to implement a well-timed relapse prevention plan. Moreover, if combined with medical care and mental support, relapse prediction can prevent relapse or extend the period that an addict stays clean. Even though relapse prediction could potentially be helpful for relapse prevention, however, it remains a very challenging task for two main reasons: (1) addicts tend to be very emotional as a result of the way drugs affect their brains, making their responses and behavior patterns inherently unpredictable~\cite{koob2009dynamics}; and (2) traditional supervised methods for relapse prediction demand a significant amount of training data if thy are to acheive reliable accuracy.


Social media is now playing an increasingly important role in addiction rehabilitation as it provides addicts with mental support during the recover process. According to a report from the National Institute on Drug Abuse (NIDA)~\cite{NIH_treatment}, opioid addiction recovery calls for not only medical treatment (such as methadone, buprenorphine and naltrexone), but also support from families, medical professionals and communities. However, family members or friends do not always know or are unsure how to help addicts to alleviate the side effects of addiction. This situation spurs addicts to turn to social media for help, posting their problems and confusion, searching for opportunities for group therapy, and seeking answers in specific forums. 
Many people who have managed to get rid of opioid dependencies are willing to share their experience with others online. For instance, on Reddit, 
some discussion groups allow users to share their opinions or experience regarding opioid addiction and rehabilitation
\textit{``I was the same way, I only got clean cuz I didn't wanna go back to jail tbh, and had to go cold turkey in rehab''}. Also, social media allows opioid addicts to express their joy and sadness, trust and disgust, anticipation and anger. These sentiments, along with addicts' online behaviors, could disclose their treatment status and thus serve as indicators for relapse prediction. We therefore sought to, capture addicts' subtle sentiments such as fear, joy, sadness or anger and then use these as target variables to predict whether an addict will relapse based on their posts on social media. 


In this paper, we present the implementation of a Generative Adversarial Net (GAN), and describe how we incorporated sentiment analysis and network influence analysis, to model the relapse prediction problem. A generative adversarial net framework~\cite{goodfellow2014generative} consists of a Generator and a Discriminator, where the generator produces data as realistically as possible and, the discriminator is then trained to determine if the data produced by the generator is real or fake. The model arrives at a global equilibrium, where the generator can generate images that are only minimally different from the real images, at a level where the discriminator is no longer able to distinguish between them. Once this has been achieved, both the generator and discriminator are considered to have learned the data distribution. Thus, if a softmax layer is applied to the output of the discriminator, the resulting GAN model can be used for relapse prediction.  
A generative adversarial net can solve the challenges mentioned above in the relapse prediction problem because:
\begin{enumerate}
    \item The sentiments are quantified and transformed for input into the generative adversarial net, thus taking into account the emotional fluctuations.
    \item GANs can learn the dependencies of different emotions from the relapse group.
    \item A generative adversarial net does not require a large amount of data since the GAN model can engage in either unsupervised or semi-supervised learning~\cite{salimans2016improved,goodfellow2016nips}.
\end{enumerate}
In this study, we gathered appropriate social media data, especially data from groups such as ``opiates", ``opiatesRecovery" and ``Drugs" on Reddit, and built a classifier to extract data about the addicts. We then quantified the emotions represented in the comments made by the addicts and transformed these sentiment scores into sentiment images that could be fed into the discriminator. Using the GAN framework, the generator was able to learn the emotion patterns using a convolution neural network while the discriminator inferred the relapse probabilities based on a neural network with a softmax layer. 
Our key contributions are:
\begin{itemize}

    \item  We designed an innovative Generative Adversarial Network to predict opioid relapse that includes a discriminator that is capable of learning the data distribution patterns of both relapsing and non-relapsing addicts.
    \item This work provides an entirely new approach to opioid relapse prediction. Classifiers are trained to identify two types of addicts: those who are struggling to get off drugs and those who enjoy the drugs and have no plans to rehabilitate. Using social media data, we seek to predict whether a recovering opioid addict will relapse or not. 
     \item We propose the concept of sentiment images that adequately capture the dynamic emotion fluctuations of addicts, as well as considering the social influences between users and the rich relationships between them.
     \item We performed a comprehensive set of of experiments using a real-world dataset that produced results that consistently outperformed those obtained using other approaches, which could potentially revolutionize the practice of opioid addiction prevention and treatment.
\end{itemize}

\section{Related Work}
We review three categories of related work as follows.

\paragraph{\textbf{Biomedical Knowledge Mining on Social Network}}
As social media are profoundly and widely intertwined with people's everyday lives, they contain more and more biomedical information about users. As a result, social media represent an increasingly important data source for discovering biomedical knowledge in applications. Bian et al. proposed the use of a Support Vector Machine to identify drug users and adverse events based on Twitter messages~\cite{bian2012towards}, while Fan et al. introduced an AutoDOA to automatically detect opioid addicts by building a Heterogeneous Information Network (HIN)~\cite{fan2017social}. Other researchers have also used supervised classification techniques to detect posts that involve medication abuse~\cite{sarker2016social}, or developed web platforms to track prescription drug abuse on social media~\cite{cameron2013predose}. All of these studies apply data analytics to social media data to detect drug addicts or detect drug abuse events, laying the foundation for further studies on topics such as opioid addict relapse prediction.
\paragraph{\textbf{Emotional States and Addiction Relapse}}
Negative emotional states, such as stress, depression, anxiety, and other emotional states are known to be related to relapse. Shiffman evaluated reports of relapse in ex-smokers, finding that most of the subjects (71\%) had negative emotions preceding their relapse, with the most common mood state being anxiety, followed by anger, frustration and depression~\cite{shiffman1982relapse}. Cummings et al. found that negative emotional states accounted for $30\%$ of all relapse~\cite{Cummings1980}. Those studies and their results indicate that emotions can indeed be useful indicators of people's status.


\paragraph{\textbf{Adversarial Neural Networks for Classification}} A number of methods for learning a classifier from unlabeled data or only partially labeled data have been proposed. Traditional solutions to this problem lie either in generative clustering methods, which try to model the data distribution directly or in discriminative clustering methods, which separate the data without modeling its distribution~\cite{springenberg2015unsupervised}.
There are also various methods for unsupervised and semi-supervised learning utilized by the neural network community. Among all the neural networks, Generative Adversarial Nets (GANs) stand out for their performance over a wide range of applications. Since their introduction by Goodfellow et al~\cite{NIPS2014_5423}, they have been used to generate images or videos~\cite{mirza2014conditional}, transform text to images~\cite{DBLP:journals/corr/ElgammalLEM17}, transform image styles~\cite{isola2017image}, generate sequences of text~\cite{yu2017seqgan}, protect medical information privacy~\cite{frid2018synthetic}, and classify images ~\cite{goodfellow2016nips}, among others. 
Springenberg proposed CatGAN that are based on an objective function that trades off mutual information between observed examples and their predicted categorical class distribution, demonstrating the robustness of the learned classifiers~\cite{springenberg2015unsupervised}. 
Papernot et al. presented a variety of new architectural features and training procedures that can be applied to GANs classifier; their proposed model was verified on MNIST, CIFAR-10 and SVHN~\cite{salimans2016improved}. These earlier studies on GAN laid the foundation for applying GAN to some interesting fields, including classification and prediction.


\section{Problem Formulation}
In this section, after introducing the formal definition of the problem, we present the solution. Our mainly focus here is building a practical and reasonable model using social media data to predict who is likely to relapse \textbf{in the next week}.
The reason why we are limiting our goal to predict relapse in the next week is based on several facts. First, relapse predictions for a shorter or a longer periods makes no sense, as $99\%$ of the recovering addicts will continue to abstain for a day since their last relapse or since beginning of the detoxification process, while more than $85\%$ will relapse within a year~\cite{sinha2011new}. Moreover, the majority of current relapse prediction research is measured in terms of weeks~\cite{bauer2001predicting,ashare2013first}, hence our focus on predicting relapse in the week ahead, utilizing it as the default setting in our experiments.

\begin{definition}[Drug Relapse]
Drug relapse is defined as the resumption of drug use after a period of abstinence. 
\end{definition}

A relapse is generally associated with younger age, heavy use before treatment, a history of injecting, and not following up with aftercare~\cite{smyth2010lapse, chalana2016predictors}. However, it is also frequently associated with extreme emotions and stressors~\cite{ford2006trauma, elster2009strong, sinha2009enhanced}. Specifically, people in rehabilitation suffer from extreme emotional swings. Because of this, this paper predicts the probability of relapsing based on emotional behaviors. 

\begin{definition}[Social Community]\label{subredit}
Each social community of an addict consists of several social groups, which is called subreddit, denoted as ${S_j}=\{{s_1^j},{s_2^j},{s_3^j},...\}$. A subreddit consists of a collection of posts, which is denoted by ${s_i^j}=\{p_{i1}^j,p_{i2}^j,p_{i3}^j,...\}$.
\end{definition}

We can conclude that the influence an addict receives is $ P=\sum\limits_{j=1}^J {\sum\limits_{i=1}^I {\sum\limits_{k=1}^K {p_{ik}^j}}}$. For an addict $a_j$, we quantify it with $<P_j,F_j,Y_j>$, where $P_j$ is the social influence received, $Y_j$ is the relapsing indicator, and $F_j$ represents the features of the addicts. In short, we combine $<P_j, F_j>$ and rewrite it as $X_j=<P_j,F_j>$.

\begin{definition}[Task]
Given a set of addicts $A=\{a_1, a_2,a_3,...\}$, and their corresponding feature variable $X_j=<P_j,F_j>$ derived from their social community ${S_j}=\{{s_1^j},{s_2^j},{s_3^j},...\}$, a task is to learn a function $f:X_j \to Y_j$, such that $Y_j \in [0,1]$, where $X_j$ and $Y_j$ are feature variable and target variable respectively.
\end{definition}

\begin{definition}[Relapsing Prediction Task]
Given a set of addicts $A=\{a_1, a_2,a_3,...\}$, and their state features ${X_j}=\{P_j,F_j\} \in R^m$, find a mapping $\Psi:R^m \to Y$, where $Y \in [0,1]$, such that the task error is minimized. Formally, we have
\begin{equation}
    \arg \mathop {\min }\limits_\Psi  \sum\limits_{j = 1}^m {||f({X_j}) - {Y_j}|{|_2}}
\end{equation}
\end{definition}


In the following part, we will discuss how to apply Generative Adversarial Networks to the relapsing prediction problem.
Given real data with distribution ${P_r}(x)$ over a feature space $X$, and expect to learn a distribution ${P_g}(x)$ that is as close as possible to ${P_r}(x)$. The closeness between two probability distribution $P_1$ and $P_2$ is defined by the divergence function $D(P_1,P_2)$. So the optimality is defined by 
\begin{equation}
  P_g^* = \arg {\inf _{{P_g}}}D({P_g},{P_r})
\end{equation}
where $P_g$ is the data distribution generated by generator, $P_r$ is the data distribution of addicts and ${\inf _{{P_g}}}D({P_g},{P_r})$ is the infimum of $D({P_g},{P_r})$.

\subsection{Generative Adversarial Networks}
Goodfellow et al. (2014), who introduced the Generative Adversarial Networks (GANs) framework, proposed it as a sort of game between a generator and a discriminator. 
A generator is trained to create fake samples that could fool the discriminator, while the discriminator is trained to distinguish between the real samples and the fakes. Thus, after each iteration, the generator gets better at creating realistic fakes, while the discriminator improves its ability to tell which are real and which fakes. The model reaches equilibrium when the generator begins to create samples that fool the discriminator. 

The GANs framework consists of two neural networks that are trained jointly. The first neural work is implemented as a generator, and it is initialized with some random noise $x_0$. The discriminator is trained using traditional supervised learning techniques. The input of the discriminator then consists of two types, real data (labeled) and fake data created by the generator. The output is scalar, simply indicating if the data is real or fake. In theory, these two opponents playing in the game are represented by two functions, each of which must be differentiable both concerning its inputs $x$ and its parameters $\theta$. The discriminator is represented by a function D that takes $x$ (a combination of real and fake data) as input and uses $\theta ^{(D)}$ as parameters. Similarly, the generator is represented by a function $G$ that takes z as input and use ${\theta ^{(G)}}$ as parameters. The objective function of the generator is to minimize the cost ${J^{(G)}}({\theta ^{(D)}},{\theta ^{(G)}})$ while controlling only ${\theta ^{(G)}}$. Similarly, the objective function of the discriminator is to minimize the cost ${J^{(D)}}({\theta ^{(D)}},{\theta ^{(G)}})$ while controlling only ${\theta ^{(D)}}$. The solution to this game is local differential Nash equilibrium, which is described by a tuple $({\theta ^{(D)}},{\theta ^{(G)}})$~\cite{ratliff2013characterization}.

\begin{algorithm}[thbp]
   Initialize ${\theta}_d$ for D and ${\theta}_G$ for G\;
   \For{each training iteration}{
        \For{K steps}{Sample m examples ${x^1,x^2,...,x^m}$ from data distribution $P_{data}(x)$\;
        Sample m noise samples ${z^1,z^2,...,z^m}$ from the prior $P_{prior}(z)$\;
        Obtaining generated data ${{\widetilde x}^1,{\widetilde x}^2,...{\widetilde x}^m}$ by using ${\widetilde x}^i=G(z^i)$\;
        Update discriminator parameters ${\theta}_d$ to maximize $\widetilde V = \frac{1}{m}\sum\limits_{i = 1}^m {\log D({x^i}) + \frac{1}{m}} \sum\limits_{i = 1}^m {\log (1 - D({{\widetilde x}^i}))}$ and  ${\theta _d} \leftarrow {\theta _d} + \eta \nabla \widetilde V({\theta _d})$\;}
        Sample another m noise samples ${z^1,z^2,...,z^m}$ from the prior $P_{prior}(z)$\;
        Update generator parameters ${\theta}_g$ to minimize $\widetilde V = \frac{1}{m}\sum\limits_{i = 1}^m {\log D({x^i}) + \frac{1}{m}} \sum\limits_{i = 1}^m {\log (1 - D({z^i}))}$ and  ${\theta _d} \leftarrow {\theta _d} - \eta \nabla \widetilde V({\theta _d})$\;
   }
\caption{Generative Adversarial Nets~\cite{NIPS2014_5423}}
\label{algorithm:Q-learning}
\end{algorithm}

\begin{figure}[hbpt]
 \centering
 \includegraphics[width=\linewidth]{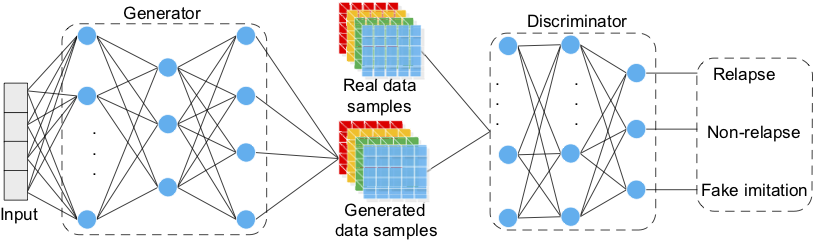}
 \caption{Generative Adversarial Net Framework.}
 \label{fig:gan}
\end{figure}

\subsection{Cost Functions}
Since the discriminators of GANs are similar to a supervised binary classifier, they have the same cost function. Suppose the discriminator is approximated by a function $D:R^n \to [0,1]$, and let $(x_i,y_i)\in (R^n,{0,1})$ be the input-label pairing for training data points, the cross-entropy of the binary discriminator is defined as
\begin{equation}
 H(G) =  - \sum\limits_{i = 1}^N {({y_i}logD({x_i})}  + (1 - {y_i})log(1 - D({x_i})))
\end{equation}
In the case of GANs, since $x_i$ only comes from two sources: either $x_i\sim p_{data}$, the true data distribution, or $x_i=G(z)$ where $z \sim p_{generator}$ (the generator's distribution) based on some input code z. The z samples might be from any distribution, but for simplicity we will specify it as $z\sim unif[0,1]$. 
Also, the best model needs precisely half of the data to come from each of these two sources. To apply this to the cross-entropy function above, we need to transform this probabilistically, after replacing the sums with expectations, the $y_i$ labels with $1/2$, and furthermore replacing the $\log (1 - D({x_i}))$ term with $\log (1 - D(z))$ under some sampled code z for the generator. We get
\begin{equation}
  J({\theta ^{(D)}}) =  - \frac{1}{2}{E_{x \sim {p_{data}}}}[\log D(x)] - \frac{1}{2}{E_{z}}[\log (1 - D(G(z)))]
\end{equation}


The cost function of the generator varies according to different specifications. First of all, in the zero-sum game, the sum of all players' costs is always zero. That is $J^{(G)}=-J^{(D)}$, and hence the cost function is
\begin{equation}
    J({\theta ^{(G)}}) =  \frac{1}{2}{E_{x \sim {p_{data}}}}[\log D(x)] + \frac{1}{2}{E_z}[\log (1 - D(G(z)))]
\end{equation}

Because $J{^{(G)}}$ is closely tied to $J{^{(D)}}$, we can summarize the entire game with a value function specifying the discriminator's payoff:
\begin{equation}
    V({\theta ^{(D)}},{\theta ^{(G)}})=-J^{(D)}({\theta ^{(G)}},{\theta ^{(D)}})
\end{equation}

The solution to a minimax game involves  minimization in an outer loop and maximization in an inner loop:
\begin{equation}
    {\theta ^{(G)*}} = \arg \mathop {\min }\limits_{{\theta ^{(G)}}} \mathop {\max }\limits_{{\theta ^{(D)}}} V({\theta ^{(D)}},{\theta ^{(G)}})
\end{equation}
The primary purpose of the generative adversarial net is for generating reliable samples according to some constraints, but it can be modified into a classifier with some machine learning techniques. Next, we will explain how to adjust it to fit for relapse prediction.
\subsection{Relapsing Prediction}
The relapse prediction model is a combination of GAN and a standard neural network classifier. For a standard classifier that classifies a data point x into one of K possible classes, it takes in x as input and outputs a K-dimensional vector of logits $\{l_1,...,l_K\}$, that can be turned into class probabilities by applying the softmax: ${p_{model}}(y = i|x) = \frac{{\exp ({l_j})}}{{\sum\limits_{k = 1}^K {{l_k}} }}$. This model is then trained by minimizing the cross-entropy between the observed labels and the model predictive distribution $p_{model}(y|x)$. Similarly, a GAN classifier is built with standard neural classifier by simply adding samples from the GAN generator to the data set. The samples are then labeled with a new `generated' class $y = K + 1$, and correspondingly increasing the dimension of the output of the classifier from K to K + 1. In this case, $p_{model}(y = K + 1 | x)$ indicates the probability that x is fake, corresponding to $1-D(x)$ in the original GAN framework. This process is given in Figure~\ref{fig:gan}. This model can also learn from the unlabeled data, that corresponds to one of the K classes of real data by maximizing $log p_{model}(y \in \{1,..., K \}|x)$~\cite{salimans2016improved}. Assuming half of the data set consists of real data and half of it is generated (this is arbitrary), the loss function for training the classifier then becomes

\begin{equation}
    \begin{split}
    {L}&=-{E_{x,y\sim p_{data}{(x,y)}}}{[log}{p_{model}}{(y|x)]}-\\ &{E_{x\sim G}}{{[log }}{p_{model}}{(y=K+ 1|x)]} \\
       & = {L_{supervised}}{  +  }{L_{unsupervised}}
    \end{split}
\end{equation}



where
\begin{equation}
    \begin{split}
        {L}_{supervised}=-{E_{{x,y}\sim {p_{data}}{(x,y)}}}log {{{p}}_{model}}{(y|x,y<K+1)}\\
        &
    \end{split}
\end{equation}

\begin{multline}
    {L_{unsupervised}}  = - {E_{{x\sim}{p_{data}}{(x)}}}{log[1 - }{p_{model}}{{(y = K + 1|x)] }}-\\ {E_{x\sim G}}{log[}{p_{model}}{(y = K + 1|x)]}
\end{multline}


Once the generative adversarial net model has been trained, we obtain a model that can be used either to predict relapse or to generate sample data. On the one hand, the discriminator has now learned the data distribution of the relapsing population and can thus be used to predict relapses. On the other hand, since the generator has learned what real relapsing data looks like, it can produce realistic fake data by imitating the real data. This process is shown in Figure~\ref{fig:gan}. 

\section{Model Development, Application and Results}

\subsection{Experimental Design}
In this section, we utilize an experimental study using real data from Reddit.com to comprehensively evaluate the performance of our proposed new GANs model. A series of complex intermediate experiments were involved including, for example, natural language processing and image processing, and these are described in turn below. Finally, the proposed method is evaluated by comparing its performance with those of alternative methods. The evaluation metrics are listed in Table~\ref{tab:measurements}.

\begin{table}[htbp]
  \centering
  \caption{Performance indices of classification}
    \begin{tabular}{ll}
    \toprule
    Indices & Description \\
    \midrule
    TP    & \# of correct classification as positive \\
    TN    & \# of correct classification as negative \\
    FP    & \# mistakenly classified as positive \\
    FN    & \# mistakenly classified as negative \\
    Precision & TP/(TP+FP) \\
    Recall & TP/(TP+FN) \\
    ACC   & (TP+TN)/(TP+TN+FP+FN) \\
    F1    & 2*Precision*Recall/(Precision+Recall) \\
    \bottomrule
    \end{tabular}%
  \label{tab:measurements}%
\end{table}%

The GANs model described here consists of a generator and a discriminator that are trained together. In this case, the generator generates data about relapsing addicts, while the discriminator evaluates whether the input data it receives is from a real relapsing addict or not. The input data of the discriminator therefore consists of a combination of real data from addicts and fake data generated by the generator. Meanwhile, the generator continues to create new data that imitates real data from relapsing addicts in an attempt to fool the discriminator. The data it generates is fed into the discriminator, along with a stream of real data from actual relapsing addicts. The goal of the discriminator is to identify the data produced by the generator as fake. The discriminator receives three categories of data: data about addicts who have relapsed, data about addicts who are struggling to stay clean and data generated by the generator. The output of the discriminator is thus a vector of probabilities with three elements that indicate the probability that the input belongs to each category. The model that emerges at the end of this process can be used both to predict relapse probabilities and generate sample data.

\begin{table}[htbp]
  \centering
  \caption{Data description}
    \begin{tabular}{lrrr}
    \toprule
    Dataset & \multicolumn{1}{l}{\# of users} & \multicolumn{1}{l}{\# of posts} & \multicolumn{1}{l}{\# of comments} \\
    \midrule
    Group\_1 (Opiates) & 1137  & 474   & - \\
    Group\_2 (Opi. Rec.) & 1621  & 456   & - \\
    Group\_3 (Drugs) & 1271  & 331   & - \\
    personal\_1 (Opiates) & 1137  & -     & 40258 \\
    personal\_2 (Opi. Rec.) & 1621  & -     & 30247 \\
    personal\_3 (Drugs) & 1271  & -     & 32403 \\
    \bottomrule
    \end{tabular}%
  \label{tab:dataset}%
\end{table}%

\subsection{Data collection}


A web crawling tool was developed using PRAW (Python Reddit API Wrapper) to collect data from Reddit.com, a social media platform open to specific groups and sessions. The dataset for this paper came primarily from three subreddits (see Definition~\ref{subredit}) on Reddit.com, namely, "Opiates", "OpiatesRecovery" and "Drugs". Within each subreddit, there are a series of discussion posts (see Definition ~\ref{subredit}). We collected the data using keyword-based method, with the keywords being a list of opioid-related drug names and its abbreviations, slangs or jargons that are related to opioids, such as `oxy (oxytocin)', `Xanax', `dope', etc. The dataset consists of two parts. The first consists of the subreddits, which include comments and interactions within specific posts. After collecting the subreddit data, the user IDs of the users who either create a post or make at least one comment are extracted to produce a list user IDs from the corresponding subreddit. The second part of the dataset consists of personal data, which was collected based on the user list extracted from the first part. Based on each user ID, comments from the previous 37 days was collected using keyword-based methods; and comments/posts from the first 30 days were then used to build sentiment images, which are explained in in more detail in Section~\ref{sentiment_image}, and posts/comments from the final 7 days used to label individual users as relapsed or not lapsed based on self-reporting. This involved manually reading the posts/comments from the last 7 days; if a user posts/comments that he/she has relapsed, then they are labeled as a relapsed addict.
In summary, we collected comments from \textbf{1,261 Reddit posts}. By extracting user id from subreddits, we also collected comments containing personal history data from \textbf{4,029 Reddit users}, including about \textbf{102,908 comments}. A summary of the dataset is given in Table~\ref{tab:dataset}.


\begin{table*}[htbp]
  \centering
  \caption{Classification results}
    \begin{tabular}{lrrrr|rrrr}
    \toprule
    \multicolumn{5}{c|}{Addict Classification} & \multicolumn{4}{c}{Recorvering Classification} \\
    \midrule
          & \multicolumn{1}{l}{Accuracy} & \multicolumn{1}{l}{Recall } & \multicolumn{1}{l}{Precision} & \multicolumn{1}{l|}{F1} & \multicolumn{1}{l}{Accuracy} & \multicolumn{1}{l}{Recall } & \multicolumn{1}{l}{Precision} & \multicolumn{1}{l}{F1} \\
    Log. Regr. & 0.8180 & 0.8277 & 0.8180 & 0.8228 & 0.7433 & 0.7834 & 0.7334 & 0.7575 \\
    KNN   & 0.7027 & 0.8144 & 0.7592 & 0.7858 & 0.7521 & 0.8121 & 0.7327 & 0.7703 \\
    SVM   & \textbf{0.9176} & \textbf{0.9429} & \textbf{0.9346} & \textbf{0.9388} & \textbf{0.8762} & \textbf{0.9290} & \textbf{0.8447} & \textbf{0.8849} \\
    \bottomrule
    \end{tabular}%
  \label{tab:classification}%
\end{table*}%

\begin{figure}[h]
	\centering
	\includegraphics[width=\linewidth]{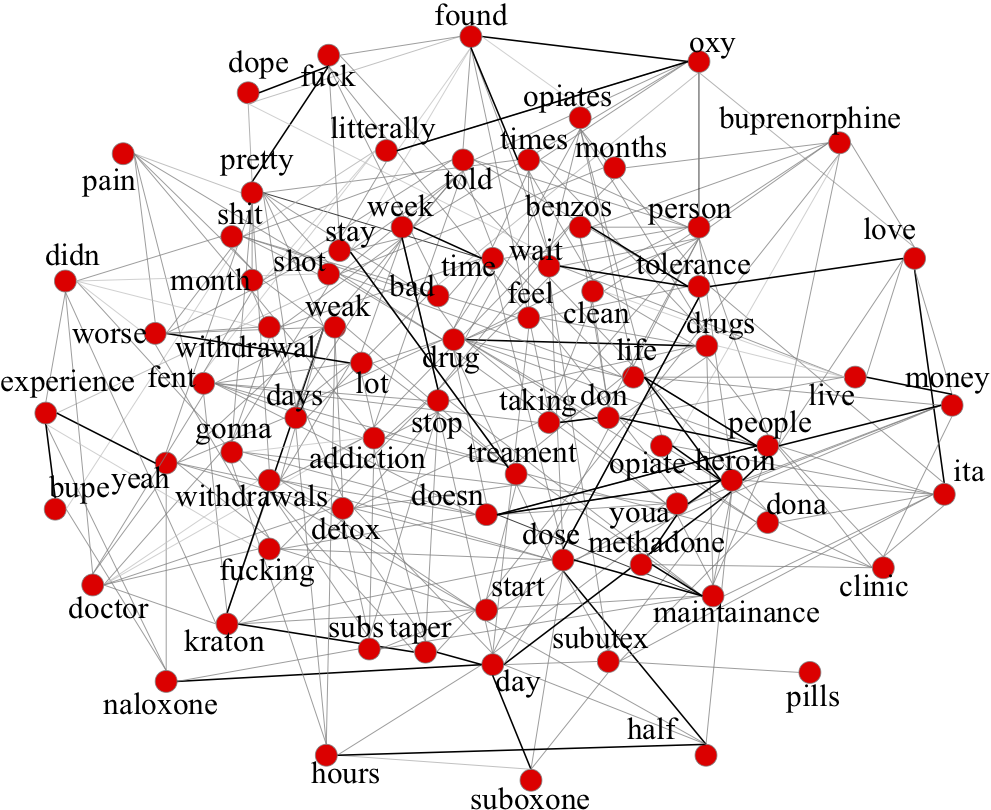}
	\caption{Word Correlation for words from group discussion of `OpiatesRecovery' with $\phi\geq 0.15$. 
	}
\label{fig:word_small_corr}
\end{figure}

\begin{figure*}[!h]
 \centering
 \includegraphics[width=\linewidth]{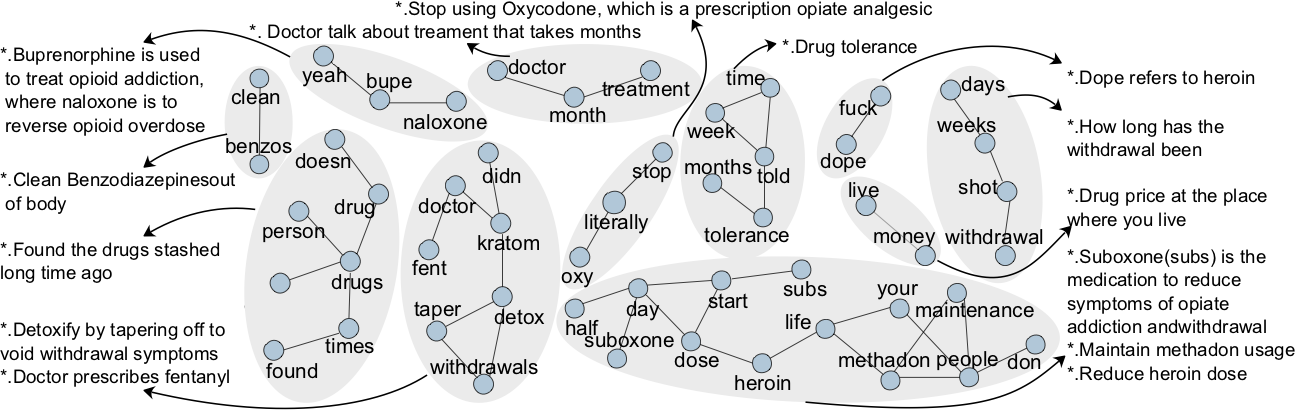}
 \caption{Word correlation and clusters with $Phi$ coefficient $\phi \geq 0.2 $. The $Phi$ value indicates the possibility of co-occurrence within a post.}
 \label{fig:word_large_corr}
\end{figure*}

\subsection{Text Co-occurrence and Correlation}

To better understand the social interactions between different users, we began by generating word correlation graphs to obtain a full picture of their conversations. Here, the textual analysis was based on the relationship between words, with the objective being to examine words that tend to immediately follow others, or that tend to co-occur within the same posts. We first tokenized comments into consecutive sequences of words. By checking how often word X was followed by word Y, we were then able to model the relationships between them using the Phi efficient metric, which is defined as
\begin{equation}
    \phi  = \frac{{{n_{11}}{n_{00}} - {n_{10}}{n_{01}}}}{{\sqrt {{n_{1.}}{n_{0.}}{n_{.0}}{n_{.1}}} }}
    \label{equation:phi}
\end{equation}
where $n_{11}$ represents the number of posts where both $X$ and $Y$ appear, $n_{00}$ is the number where neither appears, and $n_{10}$ is the cases where one appears without the other.

By gradually increasing the Phi coefficient, we manually filtered out the words that were most likely to appear with each other. After analyzing these posts, we visualized them as shown in Figure~\ref{fig:word_small_corr} and Figure~\ref{fig:word_large_corr}. By analyzing and annotating these keywords, we summarized the most frequent topics as follows:
\begin{itemize}
    \item Talking about the detoxification processes and medicines, such as `naloxone', `bupe', `doctor', and `detox'.
    \item Describing pains and prescriptions from the doctor, as described by keyword such as `pain', `doctor',`painkiller'.
    \item Discussing withdrawal symptoms.
    \item \textbf{There are two types of addicts: the first type includes the ones who are in recovery or struggling to recover, while the other type includes the ones who are not in the addiction recovery or not seeking recovery.}
\end{itemize}


\subsection{Addict Classifier}

Two classifiers were designed and implemented to filter out the research targets: an addict classifier and a recovering classifier. The first of these differentiates between addicts and non-addicts, while the second ``within addicts" classifier separates those who are in recovery and who are not seeking recovery. 
Since an opioid rehabilitation must go through the detoxification process, it is meaningless to predict whether the second type of addicts will relapse. The following research mainly focuses on the first type of addicts, who are in recovery but struggling with relapse temptations.

The addict classifier is used to classify addicts from non-addicts.
Based on users' historical posts, we employed a Support Vector Machine (SVM) classifier to determine whether a user was addicted. In most of the cases, users admitted their drug dependencies. Where we were unable to identy a user as an addict, they were labeled as a non-addict. The dataset for this classifier, which consisted of 1,000 users (419 addicts and 581 non-addicts), took four people four days to compile as it required the researchers to 
read through all the comments made by each user and label them accordingly. Of the entire labeled dataset, $70\%$ (namely, $70\%$ of the users) was used as the training dataset, and the rest was used for testing. Once the classifier had been trained, we applied it to the unlabeled dataset. Identified addicts were then fed into the next classifier to determine whether they were still enjoying using drugs or recovering. 

The second classifier was designed to identify the addicts who were recovering. As mentioned earlier, there are two types of addicts. The first type of addict enjoys drugs, and hence will almost certainly relapse, so the relapse prediction for this group makes no sense. A similar method was then applied to identify these addicts who were willing to become drug-free but struggling with the process. This took another four days to label 1,000 users (with 375 recovering users and 625 users who still enjoyed drugs). Among this labeled data, $70\%$ was used as the training dataset and the rest for testing. The performance of the two classifiers is shown in Table~\ref{tab:classification}.

\begin{figure}[htbp]
 \centering
 \includegraphics[width=0.8\linewidth]{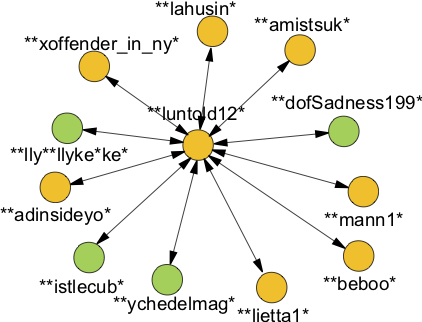}
 \caption{Interaction network for user `**luntold12*' from `Opiates' group on Reddit, where green dot represents the users who are currently clean and yellow dot represents users who are addicts. User `**luntold12*' may be influenced by various users. }
 \label{fig:network_individual}
\end{figure}

\begin{figure}[thbp]
 \centering
 \includegraphics[width=\linewidth]{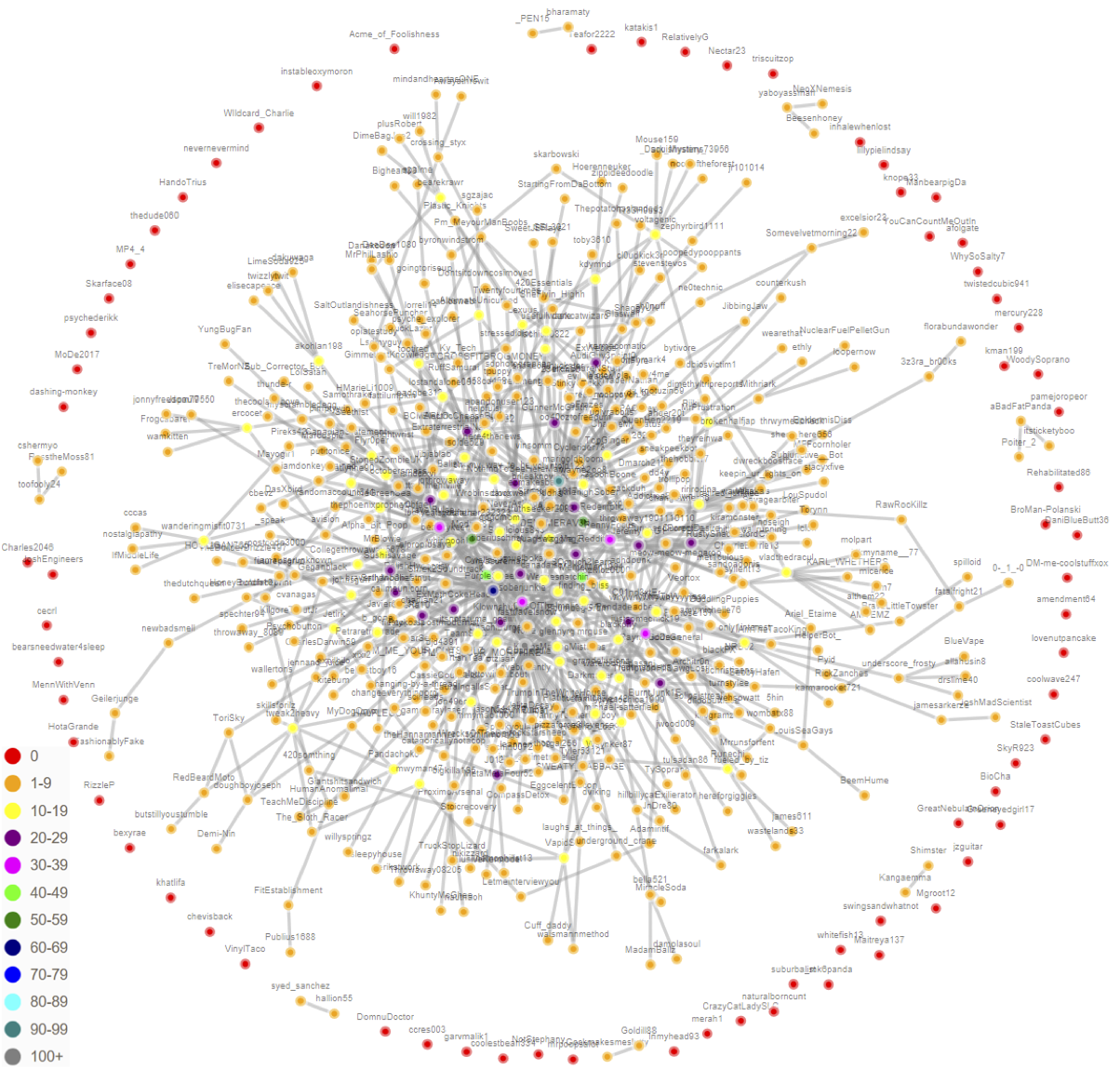}
 \caption{Network for group interactions of Reddit users from `Drugs', `Opiates' and `OpiatesRecovery'. Different colors represent the frequency of interaction in our dataset. }
 \label{fig:network_group}
\end{figure}

\begin{figure}[t]
    \centering
    \includegraphics[width=0.99\linewidth]{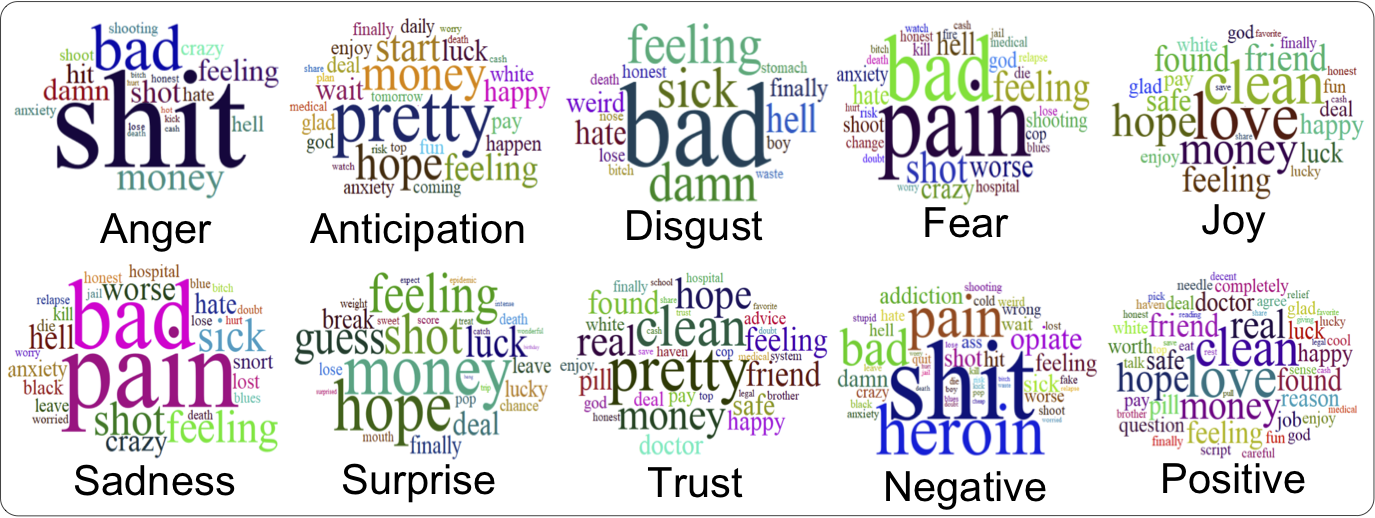}
    \caption{Word cloud for different emotions.}
    \label{fig:wordcloud}
\end{figure}
\subsection{Emotion Measure}
A word-emotion association lexicon provided by Mohammad et al.~\cite{Mohammad13} was utilized in this study to quantify the emotional content of each comment from a user. Ten categories of emotion are identified in this lexicon: \textit{anger, anticipation, disgust, fear, joy, sadness, surprise, trust, negative, positive}. Some sample words from each category are given in Figure~\ref{fig:wordcloud}. Each of these categories is associated with an emotion count variable. If a term used in a comment is listed in one lexicon category, the count variable for that category is increased by one. The process is repeated for the entire comment databse. Once the process was completed, the counts were normalized by dividing each count variable over the maximum count within the same comment to provide a relative comparison between emotions. The emotion category with the maximum value then represents the most influential or dominant emotion of this user. The smaller the value of the normalized emotional count, the less influential that emotion is. 


In particular, suppose $L_i$ is the subset of a lexicon which contains words in emotion $i$. The emotion count $emotion\_count\_i$ and normalized count $n\_emotion\_count\_i$ for the emotion $i$ are: 

\begin{equation}
    emotion\_count\_i = \sum{word_j}
\end{equation}

\vspace{-2mm}

\begin{equation}
    n\_emotion\_count\_i = \frac{emotion\_count\_i}{max(emotion\_counts)}  
\end{equation}

Where:
\vspace{-2mm}

\begin{equation}
  word_j =
  \begin{cases}
    1 & \text{if $word_j$ in $L_i$} \\
    0 & \text{otherwise}
  \end{cases}
\end{equation}


\subsection{Sentiment Image}\label{sentiment_image}
A sentiment image for an addict is defined using a sentiment matrix. After quantifying the sentiment scores for every comment, we obtained a list of sentiment scores for each addict: for each comment, the sentiment score is arranged as [Anger, anticipation, disgust, fear, joy, sadness, surprise, trust, negative, positive]. We calculated the average sentiment score for every three comments because we collected 30 comments from each user and the emotion vector has a length of 10. Next, we stacked up the average sentiment score for the thirty posts to form the sentiment image, a 10*10 matrix with each row representing the average sentiment score of three comments. The sentiment images of the addicts are then fed into the discriminator as real data from relapsing addicts.

Social influences or interactions are also considered when building the sentiment images. Just like a color image that has 3 RGB (Red, Green and Blue) channels, the sentiment image we utilized here has two channels. The first of these channels is a 10*10 matrix, which is built by stacking up ten sentiment vectors. The second channel is constructed by taking the social network influences into consideration. Specifically, if an addict starts a discussion by asking a question or describing his or her current status, other users may make comments and then the discussion creator may reply. By exchanging opinions in this way, people who exchange opinions can influence each other. For the purposes of this research, we defined social network influence as follows: 
\begin{equation}
  influence =
  \begin{cases}
    \frac{1}{n}\sum\limits_{i=1}^n{S_i}
    & \text{if n comments is replied} \\
    0 & \text{no reply}
  \end{cases}
\end{equation}

If no one replies, the discussion creator receives a 0 influence score. If dozens of users reply and the creator replies, then the creator receives an average of all the comments. For instance, in Figure~\ref{fig:network_individual} we show the social interaction with user `**luntold12*', where `*' represents a character that is hidden to protect the users' privacy. Figure~\ref{fig:network_group} shows the social interaction for users from `Opiates',`OpidatesRecovery' and `Drugs'. By analyzing the group interaction, we find that in most of the posts, the total number of comments are smaller than 10. Those participating in `Drugs' discuss illegal drug usage more, while people in `OpiatesRecovery' and `Opiates' tend to discuss how to get rid of drugs. 

The 2-channel sentiment images were then fed into the generative adversarial network to train the model.

\begin{figure}[thb]
 \centering
 \includegraphics[width=\linewidth]{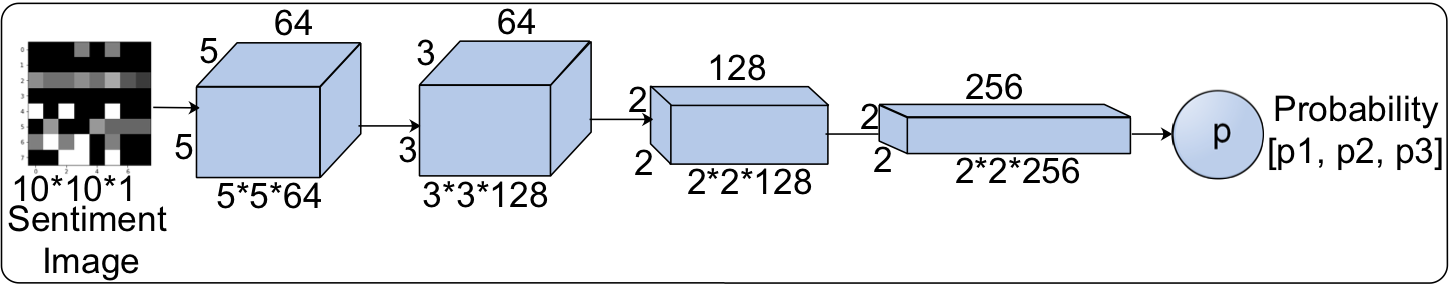}
 \caption{Input for the Discriminator and tensor flows}
 \label{fig:tensor_ds}
\end{figure}
\vspace{-4mm}
\begin{figure}[thb]
 \centering
 \includegraphics[width=\linewidth]{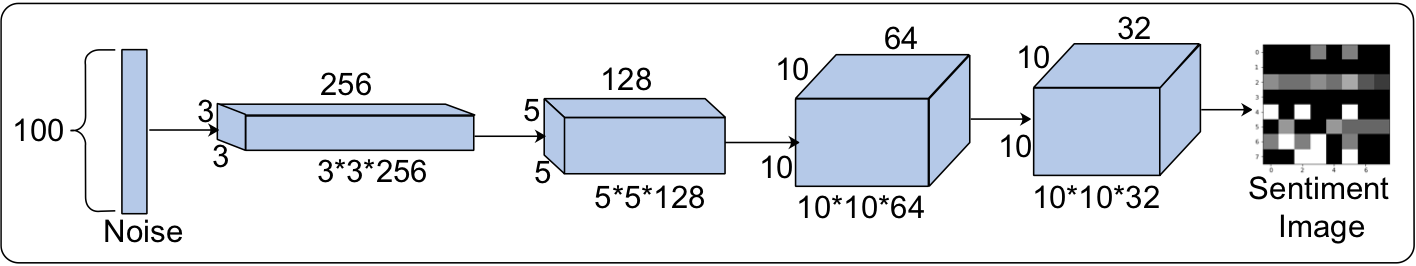}
 \caption{Input for the Generator and tensor flows}
 \label{fig:tensor_gen}
\end{figure}
\hspace{0.2cm}
\subsection{Relapse Prediction}
The relapse prediction model was modified from the discriminator of a standard generative adversarial net by applying a softmax layer to the output of a discriminator~\cite{salimans2016improved}. We trained the model for 7,000 epochs with 1,000 labeled sentiment images (consisting of 330 addicts who have managed to stay clean for a while, and 670 addicts who have relapsed). \textit{ Since no labeled dataset is available online, we labeled the data of 1,000 users} and transformed the data into images with two layers (sentiment layer and social influence layer, Channel=2). For each mini-batch, we randomly selected 128 images. For optimization we used Adam~\cite{kingma2014adam} with a learning rate=0.0001. Our implementation of the GANs model is based on Tensorflow~\cite{tensorflow2015-whitepaper}.
The data flows inside the model can be visualized as shown in Figure~\ref{fig:tensor_ds} and Figure~\ref{fig:tensor_gen}. The dimensions of the tensors are calculated as the tensor dimension calculation in convolution neural networks~\cite{convolution2018}. Figure~\ref{fig:tensor_ds} shows the input of a discriminator as a mixture of sentiment images and data from the generator, with the output being the probability that an image belongs. In Figure~\ref{fig:tensor_gen}, the input for the generator is a vector of random noise sampled from a uniform distribution, after which the data goes through several convolution layers. Finally, the generator outputs an image. 

\begin{figure}[t]
 \centering
 \includegraphics[scale=0.41]{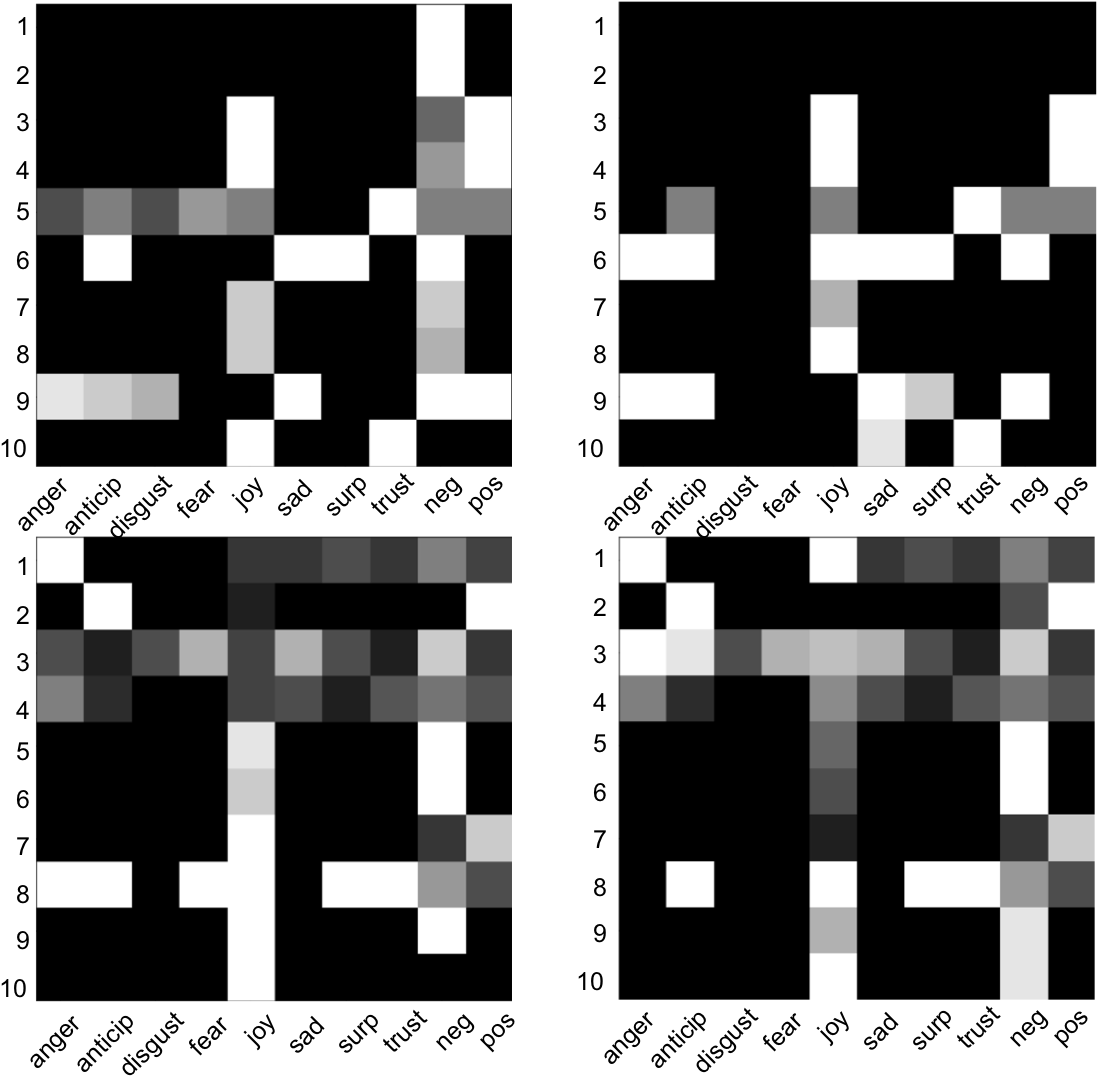}
 \caption{These sentiment images are from four addicts produced by GAN when model converged. It can be observed that the emotion `joy' and emotion `negative' are closely connected to relapse.
 }
 \label{fig:sentiment_img}
\end{figure}

\begin{figure}[t]
 \centering
 \includegraphics[width=0.9\linewidth]{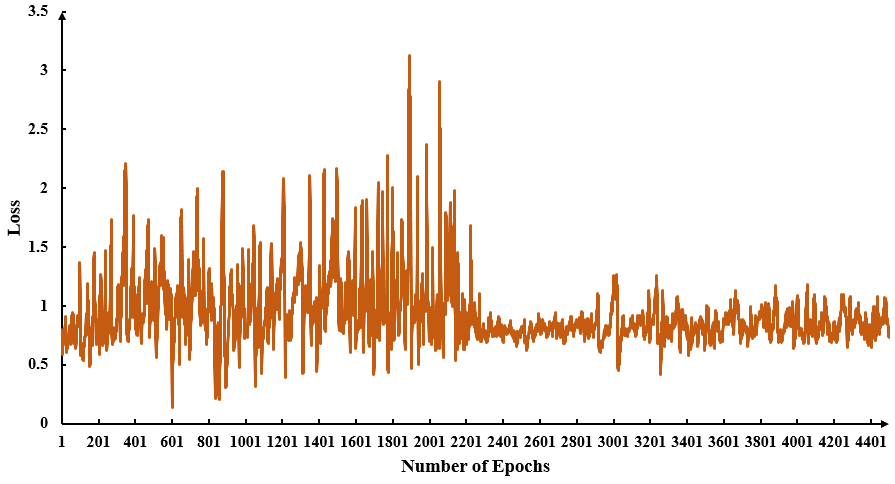}
 \vspace{-2mm}
 \caption{Generator loss changes in training.}
 \label{fig:g_loss}
\end{figure}

\begin{figure}[t]
 \centering
 \includegraphics[width=0.9\linewidth]{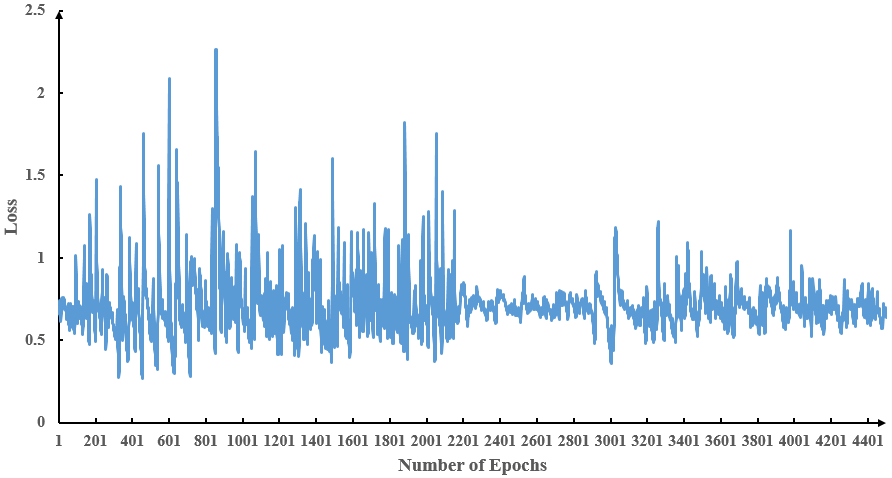}
 \vspace{-2mm}
 \caption{Discriminator loss changes in training.}
 \label{fig:d_loss}
\end{figure}

The training process is shown in Figure~\ref{fig:g_loss} and Figure~\ref{fig:d_loss}.
Figure~\ref{fig:g_loss} reveals that the generator has a random loss value while it is not converged, which indicates that the generator has not learned the patterns of input. However, the loss value becomes stable after training about 2,100 epochs, indicating that the model has converged. The whole training process for 7,000 epochs takes about 13 hours until completion. The loss value for the discriminator is shown in Figure~\ref{fig:d_loss}, once again indicating that the discriminator converged at around 2,100 epochs.

The prediction results are presented in Table~\ref{tab:prediction-table} and the new model's performance is evaluated by comparing the results with those obtained using alternative methods. The generative adversarial net achieved an accuracy of 0.8390 and an F-score of 0.8547, clearly outperforming other alternative methods such as Logistic Regression, SVM and KNN. Interestingly, the sentiment images generated by the generator show that relapse is closely connected with the emotion `joy' and the emotion `negative'. For instance, in Figure~\ref{fig:sentiment_img}, which is a set of images generated by the generator, the emotions `joy' and `negative' appear consistently in the sentiment image with non-zero values (white spaces in the figure), indicating the corresponding comments express complex feelings. The rows in the sentiment image with multiple emotions (white spaces) reveal that the user has made some lengthy comments that contain a mixture of different emotions, while the rows with fewer white spaces, signify short comments such as `so sad' and `that's bad'. 






\vspace{-2mm}


\begin{table}[h]
  \centering
  \caption{Prediction performance and evaluation}
    \begin{tabular}{rlrrrrr}
    \toprule
    \multicolumn{7}{c}{Relapse prediction with different sizes of training data} \\
          &       & 90\%  & 80\%  & 70\%  & 60\%  & 50\% \\
    \midrule
    \multicolumn{1}{l}{Log.Reg} & ACC   & 0.7150 & 0.6920 & 0.6520 & 0.6320 & 0.6020 \\
          & F1    & 0.7775 & 0.7582 & 0.7268 & 0.7065 & 0.6801 \\
    \midrule
    \multicolumn{1}{l}{SVM} & ACC   & 0.8210 & 0.7980 & 0.7680 & 0.7380 & 0.7180 \\
          & F1    & 0.8500 & 0.8412 & 0.8162 & 0.7907 & 0.7749 \\
    \midrule
    \multicolumn{1}{l}{KNN} & ACC   & 0.7120 & 0.7980 & 0.6590 & 0.6290 & 0.5990 \\
          & F1    & 0.7736 & 0.8412 & 0.7283 & 0.7020 & 0.6753 \\
    \midrule
    \multicolumn{1}{l}{GAN} & ACC   & \textbf{0.8390} & \textbf{0.8190} & \textbf{0.7840} & \textbf{0.7490} & \textbf{0.7140} \\
          & F1    & \textbf{0.8547} & \textbf{0.8491} & \textbf{0.8305} & \textbf{0.8013} & \textbf{0.7716} \\
    \bottomrule
    \end{tabular}%
  \label{tab:prediction-table}%
\end{table}%


\section{Conclusion}
In this paper, we present a new application where generative adversarial nets are utilized to predict whether a recovering opioid addict will relapse or not within the next week. 
This paper is one of the first attempts to predict opioid addict relapses using massive social media data from Reddit.com. 
By combining social media data with generative adversarial networks, our model makes predictions based on addicts' previous records and current sentiment status. Our experimental results demonstrate that the new GANs model consistently outperforms traditional prediction techniques and could thus be incorporated into a strategy for supporting practitioners working on opioid relapse prevention and addiction treatment. Unfortunately, the limitations of this study are clear: the size of manually labeled training data used in the experimental is not large enough to draw strong and consolidated conclusions. We are aware of the existence of this limitation but leave it as it was because processing and manually labeling a large amount of training data is very costly. This limitation can be fixed by investing more time to collect, process and manually label more training data. We leave the limitation for future research.



\bibliographystyle{IEEEtran}
\bibliography{reference}

\end{document}